\newread\testifexists
\def\GetIfExists #1 {\immediate\openin\testifexists=#1
    \ifeof\testifexists\immediate\closein\testifexists\else
    \immediate\closein\testifexists\input #1\fi}

\immediate\openin\testifexists=e
    \ifeof\testifexists\immediate\closein\testifexists\else
    \immediate\closein\testifexists\input e\fipsf

\tolerance=1600
\parskip=5pt
\baselineskip= 5 true mm \mathsurround=1pt
\font\smallrm=cmr8  
\font\medrm=cmr9  
\font\bigbf=cmbx12
    \def\Bbb#1{\setbox0=\hbox{$\tt #1$}  \copy0\kern-\wd0\kern .1em\copy0}
    \immediate\openin\testifexists=a
    \ifeof\testifexists\immediate\closein\testifexists\else
    \immediate\closein\testifexists\input a\fimssym.def 

\def\secbreak{\vskip7pt plus .2in \penalty-200\vskip -2pt plus -.1in}
\def\ref#1{${\,}^{\hbox{\smallrm #1}}$}
   \def\newsect#1{\secbreak\noindent{\bf #1}\medskip}
   
\def\hugeskip{\vskip12mm plus 3mm}
\def\Narrower{\par\narrower\noindent}   
\def\Endnarrower{\par\leftskip=0pt \rightskip=0pt}

\def\cl{\centerline}    
\def\ni{\noindent}      \def\pa{\partial}   \def\dd{{\rm d}}
            \def\ket{\rangle}

      \def\D{\Delta}

         \def\j{\psi}

\def\HH{{\cal H}}

\def\fn#1{\ifcase\noteno\def\fnchr{*}\or\def\fnchr{\dagger}\or\def
    \fnchr{\ddagger}\or\def\fnchr{\medrm\S}\or\def\fnchr{\|}\or\def
    \fnchr{\medrm\P}\fi\footnote{$^{\fnchr}$}
    {\scrunch#1\toe}\ifnum\noteno>4\global\advance\noteno by-6\fi
    \global\advance\noteno by 1}
    \def\scrunch{\baselineskip=11 pt \medrm}
    \def\toe{\vphantom{$p_\big($}}
    \newcount\noteno

\def\ffract#1#2{\raise .35 em\hbox{$\scriptstyle#1$}\kern-.25em/
    \kern-.2em\lower .22 em\hbox{$\scriptstyle#2$}}

\def\part#1#2{{\partial#1\over\partial#2}}

\def\bbf#1{\setbox0=\hbox{$#1$} \kern-.025em\copy0\kern-\wd0
    \kern.05em\copy0\kern-\wd0 \kern-.025em\raise.0433em\box0}

\def\deff{\ {\buildrel{\rm def}\over{=}}\ }

\def\Gbar{\raise.13em\hbox{--}\kern-.35em G}

{\ }\vglue 1truecm \rightline{SPIN-2001/09}
\rightline{ITP-UU-01/15} \rightline{hep-th/0104219} \hugeskip
\cl{\bigbf  HOW DOES GOD PLAY DICE
?}\medskip\cl{\bigbf(PRE-)DETERMINISM AT THE PLANCK SCALE}
\bigskip\cl{An Essay in honour of John S. Bell} \hugeskip

\cl{Gerard 't Hooft }
\bigskip
\cl{Institute for Theoretical Physics} \cl{University of Utrecht,
Princetonplein 5} \cl{3584 CC Utrecht, the Netherlands}
\smallskip
\cl{and}
\smallskip
\cl{Spinoza Institute} \cl{Postbox 80.195} \cl{3508 TD Utrecht,
the Netherlands}
\smallskip\cl{e-mail: \tt g.thooft@phys.uu.nl}
\cl{internet: \tt http://www.phys.uu.nl/\~{}thooft/ } \hugeskip
\ni{\bf Summary}\Narrower
 In deterministic theories, one can start from a set of
 ontological states to formulate the dynamical laws, but these may
 not be directly observable. Observable are only equivalence
 classes of states, and these will span a basis of ``beables", to
 be promoted to an orthonormal basis of Hilbert Space. After
 transforming this basis to a more conventional basis, a theory
 may result that is fundamentally quantum mechanical. It is
 conjectured that the quantum laws of the real world may be
 understood from exactly such a procedure. \Endnarrower \bigskip
\newsect{Dice}
Quantum Mechanics works. It first served as a successful doctrine
for describing the dynamical laws for electrons inside atoms.
Subsequently, it was realized that quantum mechanics must be more
universally valid, so it should also apply to the atomic nuclei
and to the elementary particles out of which these nuclei are
composed.

Including the requirements of special relativity has been a major
challenge. At first sight, it seemed that quantum mechanics was to
undergo substantial revisions in order to meet this challenge. The
words "second quantization" underlined the fact that
Schr\"odinger's equation for single particles was to be replaced
by the equations for quantized fields.

But the fundamental starting points of quantum  mechanics were not
seriously altered by the advent of quantum field theory. Rather,
they were generalized and deepened. Non-Abelian gauge theories
---~essential building blocks of the Standard Model~--- added a
further complication: the emergence of ghost particles, required
for achieving unambiguously renormalized amplitudes, but again,
one has to conclude that the fundamental dogmas of quantum
mechanics not only survived, but also served as a backbone for
these theories. In accordance with these rules, the quantized
field theory must provide for the scattering matrix ($S$-matrix)
amplitudes, and the scattering probabilities should follow from
these. Should we depart from the usual ``Copenhagen" rules for
interpreting the amplitudes, then we would not even know how to
start formulating consistent theories.

For the last couple of decades, theoreticians have been
considering the challenge to include the one remaining theoretical
requirement: general relativity, {\it i.e.}, the laws of gravity.
The problem appears to be a beautiful one, since now the dynamics
of the curvature in space and time must be submitted to the laws
of quantum mechanics. The virtual contributions to the amplitudes
due to space-time curvature diverge at the small-distance end,
just like Fermi's earliest models for the weak force used to do.
In the latter case, the divergences were successfully tamed by the
introduction of the Standard Model. Naturally, one expects similar
solutions to the problem at hand, and indeed, string theories and
their successors are claimed to come close to bringing just such a
solution.

But, they have not done so yet, and upon closer inspection one
finds that there really are reasons to be skeptical. Not only are
there numerous technical difficulties ---~the identification of
the ground state, our lack of understanding the supersymmetry
breaking mechanism, the smallness of the cosmological
constant~---, there are also more fundamental ones, which are
difficulties that have little to do with the fact that we are
dealing with strings, $D$-branes, or what not. Rather, they have
to do with the fact that we are attempting to apply quantum
mechanics to dynamical laws that should be the ultimate driving
forces of the entire Universe; one will be forced to consider
statistical ensembles of universes, and such notions will be much
more questionable than the notion of an ensemble of experiments
inside one universe. It should be kept in mind that one will never
be able to do experiments in more than one universe, and that the
`averaged value' of a quantity measured in different universes
cannot be checked against any theory.

Quantum fluctuations of the space-time metric at distances at or
below the Planck scale diverge, and the implications of this may
well be deeper than the need to re-sum some divergent perturbation
series. Under these circumstances, causality and unitarity may
become impossible to keep under control, regarding the
difficulties in establishing time ordering. It should be clear
that, at this point, the most serious difficulties may be
associated to our insistence to keep the quantum rules unchanged.

It is illustrative, in this respect, to study a beautiful toy
model for a universe: point particles gravitating in a 2+1
dimensional universe\ref{1}, while obeying Einstein's equations.
The classical equations are exactly soluble in 2+1 dimensional
space-times, during finite time intervals, but they become chaotic
near a big bang or a big crunch. Poisson brackets can be written
down for this system, but replacing these by commutators fails,
exactly because the notion of a Hilbert space becomes problematic
when it refers to entire universes.

It is here that I would like to advocate a different approach. We
should not forget that quantum mechanics does not really describe
what kind of dynamical phenomena are actually going on, but rather
gives us probabilistic results. To me, it seems extremely
plausible that {\it any\/} reasonable theory for the dynamics at
the Planck scale would lead to processes that are so complicated
to describe, that one should expect {\it apparently stochastic\/}
fluctuations in any approximation theory describing the effects of
all of this at much larger scales. It seems quite reasonable first
to try a classical, deterministic theory for the Planck domain.
One might speculate then that what we call quantum mechanics
today, may be nothing else than an ingenious technique to handle
this dynamics statistically.

Of course I am aware of the numerous studies regarding the
difficulties in devising hidden variable theories for quantum
mechanics. Deterministic theories appear to lead to the famous
Bell inequalities\ref2, and the Einstein-Rosen-Podolsky
paradox\ref3. There are various possible reasons nevertheless to
continue along this avenue. Generally speaking, we could take the
attitude that every ``no-go theorem" comes with some small-print,
that is, certain conditions and assumptions that are considered
totally natural and reasonable by the authors, but which may be
violated in the real world. Certainly, physics at the Planck scale
will be quite alien to us, and therefore, expecting some or
several of the ``natural" looking conditions to be violated is not
so objectionable. More specifically, one might try to identify
some of such conditions. One example might be the following:
turning some apparatus at will, in order to measure either the
$x$- or the $y$-component of an electron's spin, requires
invariance of the device under rotations, allowing the detector to
rotate independently of the rest of the scenery. This is unlikely
to be totally admissible in terms of Planck scale variables.

Another attitude could be to assume that deterministic approaches
might only succeed partially but not completely. Locally
deterministic approaches might become useful for the
identification of local dynamical rules, but maybe some quantum
mechanical constraint will have to be added at a later stage, thus
restoring the quantum mechanical nature of the entire description.
In some of the calculations shown below, it can be seen that such
constraints can exist, since already at an early stage we do turn
to Hilbert space techniques, and there, one may well introduce
fully quantum mechanical projection operators that only slightly
modify the dynamics.

Let us simply keep an open mind, and just see where a
deterministic approach leads us. Whether or not to call my most
recent results promising, the reader may decide for him- or
herself.

The most logical domain of physics where one may expect Quantum
Mechanics to become replaceable by a more deterministic scenario
is the Planck scale. One reason for this expectation was already
mentioned: the unwieldy strangeness of the world of stringlike
phenomena which might invalidate some of the assumptions made, be
it consciously or tacitly, when arguments using the Bell
inequalities are applied to prove hidden variables to be
impossible or at least non local. Another reason for taking the
Planck regime is the emergence of black holes there. On the one
hand, a conventional quantum mechanical description of black holes
appears to contradict General Relativity; on the other hand, a
classical (read: deterministic) description of black holes implies
the feature of information loss. Information loss may indeed be
admitted in deterministic theories, and, as we shall argue
further, information loss may be the key ingredient that could
turn a local, deterministic theory into a quantum theory.

Nevertheless, one could observe that reference to the
gravitational force is not truly necessary for our considerations.
One might hope that the search for a theory behind Quantum
Mechanics could provide for further costraints on the apparently
arbitrary constants of Nature in the Standard Model, long before
the Planck scale is reached, but I am not counting on such a
miracle.

My starting point is basically very simple. The dividing line
between quantum physics and classical physics is more subtle than
usually advertised. Hilbert space techniques, as an aid to discuss
statistical features, can be introduced for deterministic
systems\ref{4---7} just like in notoriously ``quantum mechanical"
theories. A deterministic system is characterized by a set of
evolution equations that tell unambiguously how it evolves in
time, given its initial configuration. The definition of time does
not need to be very strict. One might have a continuous time
variable or discretized  time, or time might be defined in terms
of Cauchy surfaces in a general relativistic setting. There is one
important condition that must be met by the time variable,
however: it must be on a real line (possibly with a beginning
and/or an end to it). Time is not allowed to be cyclic. If closed
time-like trajectories would exist, this would lead into clashes
and our theory would no longer be unambiguous. Closed time-like
loops, popular in some versions of gravity theories, will be
excluded.

First, we assume that the evolution is time-reversible (an
assumption that may later be relaxed). We attach a basis element
of a Hilbert space to every possible configuration of the Universe
at a given time $t$, so the ensemble of all possible
configurations spans a Hilbert space.  If now time is discrete,
then one step in time is associated to a permutation operator in
the space of states, and it corresponds to a unitary evolution
operator $U(\D t)$ in Hilbert space, where $\D t$ is the time
quantum. In each row and on each column of $U$ there is one entry
equal to 1 and the other entries are 0.

Subsequently, one may define an operator $H$ such that $$U(\D
t)=\exp(-iH\D t)\,.\eqno(1)$$ This operator is not unambiguously
defined; one may freely add or subtract multiples of $2\pi/\D t$
to its eigenvalues.

If time is continuous, we have equations in a space of coordinates
$\{q^i\}$, of the form $${\dd\over\dd t}q^i=f^i({\vec
q})\,,\eqno(2)$$ where $f^i$ may be any kind of functions of the
coordinates $q^i$. We limit ourselves to first-order differential
equations because, of course, higher order equations can be
reduced to first order ones by enlarging the space of independent
variables $q^i$. In our Hilbert space, this equation is the
Schr\"odinger equation corresponding to the Hamiltonian $$H=\sum_i
p^if^i({\vec q})+g^i({\vec q})\,,\eqno(3)$$ where $$p_i\deff
-i{\pa\over\pa q^i}\,,\eqno(4)$$ and $g^i$ is an arbitrary
function. Note that $p_i$ are the true momentum operators in the
quantum mechanical sense, even though we are discussing
deterministic theories. The point here is, that the Hamiltonian
(3) is linear in the momenta, as opposed to the usual quadratic
expressions in `real' quantum mechanics.

Using either the Hamiltonian defined in (1) or the one in (3), one
may use all conventional prescriptions from Quantum Mechanics and
calculate the evolution of any `wave function'. The absolute
squares of its coefficients may always be interpreted as
probabilities. It is important to observe that in these systems
the `collapse of the wave function' means absolutely nothing;
there are probability distributions, and as soon as one measures
something, our knowledge concerning the state collapses towards
only one of various alternative possibilities, and that is all
there is to it.

But of course our systems are not truly quantum mechanical. With
our choices of the Hamiltonian, wave functions do not `spread';
there seem to be no interference effects, and it seems to be
difficult to create `coherent states'. The question is, can one
make a link between the deterministic systems described above and
theories that look more like genuine quantum mechanical systems?

Let us first introduce some definitions. The basis elements of our
Hilbert space refer to states the `universe can really be in'. We
call these states `ontological states'. They may be realized or
not realized, and, at least in our deterministic models, there is
no possibility in between, although in our analysis we shall often
take our refuge into probability concepts.

An operator that measures which of these ontological states we are
in, will be called a {\it beable\/}, \`a la John Bell. Beables
form a set of operators that are defined at all times, and they
all commute with one another. An operator that replaces an
ontological state by another ontological state will be called a
{\it changeable\/}. Changeables may also add any kind of phase
factors to the states. All operators presently known, and used in
the Standard Model, must be changeables. Changeables do in general
not commute, but they can sometimes be diagonalised. Physicists
living in a deterministic world may be unable to distinguish
beables from changeables.

At first sight, our deterministic models may seem to be trivial,
and one may suspect that there is a wide gap between these and
truly quantum mechanical models.

Just like in `true quantum mechanics', we have a Hilbert space and
a Hamiltonian. Where exactly is the gap between deterministic and
`truly quantum mechanical' theories? A very important distinction
appears to be that the Hamiltonian (3) is not bounded from below.
There is no ground state. In ordinary quantum mechanics we do have
a ground state. But here also one might have to look at the
gravitational force. When gravitational collapse is admitted, and
protons are allowed to decay, states usually considered to be
ground states are not truly ground states. There may be an alley
here. Conversely, one might consider the Hamiltonian (3) and
impose constraints. In some models, this Hamiltonian can be seen
to be the difference between two commuting, positive
operators\ref6: $$H=H_1-H_2\quad;\qquad[H_1,H_2]=0\ .\eqno(5)$$

Imposing the constraint$$H_2|\j\ket=0\,,\eqno(6)$$ on our states
$|\j\ket$, leaves us the Hamiltonian $H_1$, which may be bounded
from below. Are there other manipulations that can be performed to
turn our system into one where the Hamiltonian does have a ground
state?

If time is discrete, the Hamiltonian is bounded, so of course
there is a ground state. Here, however, the eigenvalues are
well-defined apart from multiples of $2\pi/\D t$, so here the
problem is that the choice of the ground state appears to be
arbitrary. If our world could be related to a deterministic
scenario at the Planck scale, then why does the vacuum state of
our world appear to be unique, whereas it is either arbitrary or
ill-defined in our deterministic models?

The answer to this may be that one possible refinement has not yet
been mentioned. In the beginning of our formalism, we assumed
that, in the deterministic model, the evolution law is
time-reversible. If time is discrete, this needs not be the case.
Two different configurations might evolve into the same final
state. We call this `information loss', or `dissipation'. At first
sight, the introduction of information loss appears to be a
disaster. We now no longer have a unitary evolution matrix, so we
cannot define a Hamiltonian at all. However, the damage can be
repaired. In principle, we could decide to `remove' from
consideration all those states that now cannot be reached at all
from any configuration in the distant past. Then time
reversibility might be restored, and we are back in the previous
situation. In practice, however, this is infeasible. It may be too
difficult to identify the states without a past (``Gardens of
Eden"). Another avenue may be the following.  Instead of single
states, we now use the concept of `equivalence classes of states'.
Two states are equivalent iff within some finite time period they
both evolve into the same final state. A basis element of Hilbert
space is defined to be such an equivalence class. The time
evolution of equivalence classes is time-reversible by
construction.

How can the introduction of equivalence classes affect the ground
state problem? The Hamiltonian is still the one defined in
Eq.~(1), and its ground state is still ill-defined. However, the
notion of {\it locality\/} is severely affected. Locally, one
cannot distinguish the equivalence classes, so it may appear to be
impossible to define local operators. Notice, that this resembles
the situation in gauge theories, where gauge-invariance causes
complications in defining local observables (local operators have
a high dimensionality). In string theories, local observables are
even more difficult to define.

Our ontological states may still evolve according to a completely
local law, but the equivalence classes are not locally
well-defined. Therefore, our theory may have the property that
information cannot spread faster than a certain velocity, say the
velocity of light, but its quantum states cannot be characterized
locally. We do expect that the Hamiltonian $H$ of such a theory
can be expressed as the integral over a Hamiltonian density
$\HH({\vec x})$, with $$[\HH({\vec x}),\HH({\vec
x'})]=0\quad\hbox{if}\quad {\vec x\ne \vec x'}\quad;\qquad
H\deff\int\HH({\vec x})\dd{\vec x}\,.\eqno(7)$$

The effects of introducing information loss and the equivalence
classes may be drastic. The classes may be very large, so that the
dimensionality of Hilbert space may be greatly reduced. For
instance, in the case of black holes, the total number of quantum
states is known to grow exponentially, but only with the area of
the horizon, and not its volume ---~the well-known holographic
principle. Clearly, quantum states are related to ontological
states after applying sequences of projection operators. It is
these projection operators that may be responsible for curtailing
the Hamiltonian as well. There was arbitrariness in the definition
of $H$, but insisting on equations such as (7), we might find that
$\HH({\vec x})$ does have a natural lower bound, so that this $H$
has one as well. We suspect that this is how our Hamiltonian
receives a ground state.

In this light, one may regard our newest publication\ref7. In
there, we show that free bosons are in fact deterministic, but
only if one restricts oneself to those observables which are
invariant under a specially chosen group of gauge transformations.
One may either consider these gauge transformations as an
indelible ingredient of our theory, or one might say that these
are representing transitions about which information gets lost. It
is then the equivalence classes that possess this gauge
invariance, not the ontological degrees of freedom.

This is an example where we can see our proposal work. Besides our
deterministic model leading to second quantized free bosons of any
mass, another deterministic model could be constructed whose
Hamiltonian turns out to be that of non-interacting massless
fermions. These fermions appear to behave as flat sheets rather
than particles. Here also, one may suspect that these sheets
represent equivalence classes rather than ontological variables.
Both for the fermions and the bosons, the Hamiltonian could be
made bounded from below.  As yet the models we could produce do
not seem to be physically very realistic, let alone impressive,
but they are to be taken seriously. The search for more models is
under way. It is comforting to realize that superstring theory
makes full use of freely propagating bosons and fermions along the
string world sheet. Since the construction of deterministic
quantum theories works particularly well for free bosons and
massless fermions, these theories are ideally suited for being
incorporated into the picture advocated here.

We think the above gives some idea about the approach that I wish
to advertise. Quantum mechanical features can be obtained from
completely deterministic scenarios, in such a way that the Born
interpretation of the wave function as a probability amplitude is
completely respected. Of course, we expect the reader to be
skeptical. A number of issues should be confronted in order to
become more convincing. Most of these have to do with `quantum
weirdness'. Quantum mechanics has given us so many surprising and
counter intuitive features of the submicroscopic world that many
have come to conclude that it will be forever impossible to
construct deterministic, or `ontological' theories. Let me begin
with the relatively easy issues.

--- {\it Quantum interference.} The prototype experiment is the
electron two-slit experiment. If an electron is allowed to go only
through one of the two slits, a relatively smooth pattern of
electrons is detected on the screen behind the slits. If the
electron is given the choice between two slits, an interference
pattern arises. There may be constructive or destructive
interference.

In theories of the type proposed here, this is not at all a
forbidden phenomenon. The operators that characterize the presence
of an electron apparently cannot be beables, because the operator
that detects an electron at one of the slits does not commute with
the operator that detects an electron on the screen behind the
slits. Electrons are not the ontological variables of the
deterministic system. The fields $\j$ introduced by physicists to
describe electrons are changeables, to be constructed from the
primordial variables of the theory.

The same holds, in a more obvious way, for the spin of a particle.
Spin is the eigenvalue of a rotation operator. Surely, rotations
replace ontological states by different states. The operators
measuring spins are changeables.

--- {\it Quantum coherence and entanglement.} I am often asked
how, in deterministic theories, coherent and/or entangled quantum
states can arise. How they can arise out of inconspicuous initial
states, I do not know, but it should be stressed that the notion
of states is here as in genuine quantum mechanics. All quantum
states may be considered. We may prepare any initial state we
like, and ask how it evolves with time. It is important to
emphasize that our description of {\it states\/} is exactly as in
genuine quantum mechanics. We allow the use of operators with
non-trivial commutation rules. These states merely describe our
choice of the probablities of the various possible configurations
at any given time. So we definitely allow these probabilities to
be as in a state with `quantum coherence'. Where our theories
differ from conventional quantum mechanics is our choice of the
{\it dynamics}. The dynamical rules are deterministic, but only
after one has been able to identify what the basis of Hilbert
space is that corresponds to the ontological states, or the
equivalence classes (which are also ontological). The description
of this basis may be deeply hidden in the Planckian regime.

--- {\it The Einstein-Rosen-Podolsky paradox\/} and {\it the
violation of the Bell inequalities.} This is surely the most
difficult aspect to be addressed, and a completely satisfactory
response has not yet been given. Part of an answer can perhaps be
found in the observations made above: we do admit non-commuting
operators and states in which some of these are diagonalized. But
this does not explain why and how experiments can be performed
that defy any description in terms of local, ontological
observables. We must conclude that the observables commonly used
are not ontological. They are not beables. This however leads to
the question: how can it be that, at any given time, an
experimenter can choose to measure either the $x-$ or the $y-$, or
the $z-$ component of a spin, as if these {\it were\/} beables?

Before attempting to answer this question, we should decide about
a related topic: {\it is a registration of an experimental result,
fixed in a classical way, a beable?} If the motion of planets is
partly determined by haphazard events with a quantum mechanical
origin (particularly relevant if the motion is a chaotic one, such
as most Newtonian systems), are the coordinates of planets, or at
least their first 20 or 30 decimal places, beables?

The answer here is most probably yes. So the previous question
remains: how can changeables turn into beables in the course of a
measurement?

My suspicion is that the relation between changeables and beables
at the atomic scale is extremely complex. It is already fairly
complicated to express the $x$ variable and the $p$ variable of a
harmonic oscillator in terms of its beables, as described in
Ref\ref7. In more realistic theories that include interactions, it
might involve the changeables and beables of the experimenter
himself, when the decision is made which component of the spin to
measure. This is what Bell called {\it predeterminism}. The
experimenter's decisions also follow from Standard Model
interactions in his brain. Predeterminism is here defined by the
assumption that the experimenter's `free will' in deciding what to
measure (such as his choice to measure the $x$- or the
$y$-component of an electron's spin), is in fact limited by
deterministic laws, hence not free at all, so that it must be
included in a deterministic interpretation of Quantum Mechanics.
Most physicists dismiss predeterminism as a likely resolution of
the EPR paradox. I am not so certain about this. The entire
combination, experimenter/experiment, should be understood by
performing a renormalization group transformation from the Planck
scale to the atomic scale. But exactly how all of this should be
worked out is far from understood.

On the other hand, it also remains to be seen whether the EPR
paradox and Bell's inequalities are real contradictions at all, or
whether they can simply be ignored. If our models work, they
explain quantum mechanics, and it could be that the apparent
violation of Bell's inequalities will be seen as a natural
consequence. Although in any case, more understanding of this
situation is desired, it is legitimate to postpone answering these
questions until more sophisticated models have been found. The
existence of unanswered questions of this sort does not invalidate
our approach, so we continue research in this direction.

\newsect{References}

\item{1.}A. Staruszkiewicz, Acta Phys. Polon. {\bf 24} (1963) 734;
S. Giddings, J. Abbott and K. Kuchar, Gen. Rel. and Grav. {\bf 16}
(1984)  751;  S. Deser, R. Jackiw and G. 't Hooft , Ann. Phys.
{\bf 152} (1984) 220; S. Deser and R. Jackiw, Comm. Math. Phys.
{\bf 118} (1988) 495; G. 't Hooft, Class. Quantum Grav. 10 (1993)
1023-1038.
\item{2.} J.S.~Bell, {\it Physica \bf 1}, 195 (1964).
\item{3.} A. Einstein, B. Podolsky and N. Rosen, {\it Phys. Rev.} {\bf  47} (1935)
777.

\item{4.} G.~'t~Hooft, {\it Found.~Phys.~Lett. \bf 10} (1997) 105; quant-ph/9612018.

\item{5.} G.~'t~Hooft, {\it Class.~Quant.~Grav. \bf 16} (1999) 3263; SPIN-1999/07,
gr-qc/9903084.

\item{6.} G.~'t~Hooft, ``Determinism and Dissipation in Quantum Gravity",
presented at {\it Basics and Highlights in Fundamental Physics},
Erice, August 1999, SPIN-2000/07, hep-th/0003005.

\item{7.}  G.~'t~Hooft, ``Determinism in Free Bosons", SPIN-2001/07, hep-th/0104080.

\bye